\documentstyle[12pt]{article} \begin{document} 
\title{Quark model and strange baryon production in heavy ion collisions }

\author{A.Bialas \\ M.Smoluchowski Institute of Physics \\Jagellonian
University, Cracow\thanks{Address: Reymonta 4, 30-059 Krakow, Poland;
e-mail:bialas@thp1.if.uj.edu.pl}}
\maketitle

\begin{abstract}

It is pointed out that the recent data on strange baryon and antibaryon
production in Pb-Pb collisions at 159 GeV/c agree well with the
hypothesis of an intermediate state of quasi-free and randomly
distributed constitutent quarks and antiquarks. Also the S-S data are
consistent with this hypothesis. The p-Pb data follow a different
pattern.

\end{abstract}

Recently, rather precise  data on strange baryon and antibaryon  production
 in the central rapidity region of Pb-Pb and p-Pb collisions
 were presented by the WA97 collaboration \cite{97}. 
In this note I would like to point out that these Pb-Pb data  agree
rather well with a simple quark-counting rule whereas the p-Pb data follow
a different pattern. This observation implies that the quark degrees of
freedom are much more  relevant in collisions of two heavy nuclei than
in "elementary" hadronic interactions. It thus supports the
interpretation of the data on strangeness production in Pb-Pb collisions
as an evidence for creation of the quark-gluon plasma  \cite{rf,gz}. 

Our argument is an application of the old idea proposed first by
Rafelski \cite{rf2}. Considering a system of partons in thermodynamic
equilibrium (i.e. quark-gluon plasma), he observed that strange (and
antistrange) particle abundances must satisfy a host of simple
relations. Below we consider some of these relations which have a virtue
of being rather general, independent of the  assumption of
thermal equilibrium but -on the other hand- sensitive to the quark
degrees of freedom.

To explain the argument, let us formulate the quark counting rule we are
talking about. We simply assume that probability of creation of a baryon
(or an antibaryon) with a given quark content is proportional to the
probability that three quarks (or antiquarks) with appropriate quantum
numbers happen to meet at a certain region of phase-space - necessary
for the binding to take place. Assuming furthermore that the quarks are
uncorrelated\footnote{Both these assumptions are valid in thermal
equilibrium. The inverse is not true, however. }, we obtain the
following relative probabilities:

\begin{equation}
p = \omega_p\; q^3 ;\;\;\;\; \Lambda/\Sigma^0=\omega_{\Lambda}\;q^2s;\;\;\;\;
 \Xi =\omega_{\Xi} \;qs^2
;\;\;\;\; \Omega =\omega_{\Omega}\; s^3   \label{1}
\end{equation}
where $q$ and $s$ are relative probabilities to find a light quark and a
strange quark in the suitable phase-space region.
 Analogous formulae are valid for antibaryons. 
$\omega_i$ are  the 
proportionality factors taking into account
 the effects of resonance structure and of the binding energy
in formation of various baryons.
These factors, generally  different for different baryons, are rather 
difficult to calculate and therefore  
the comparison of Eqs (\ref{1}) with experiment is rather involved and
depends on further assumptions \cite{rf,be}.

 One may observe, however, that {\it these $\omega$-factors
 are identical for a baryon and the
 corresponding antibaryon}. Consequently,   if one
considers only the {\it ratios} of the antibaryon to baryon rates, the
$\omega$-factors cancel \cite{rf2,let} and the discussion becomes much
simpler. This is what we are going to do. We thus have
\begin{equation}
\frac{\bar{p}}{p} = \frac{\bar{q}^3}{q^3}    \label{2}
\end{equation}
and 
\begin{equation}
\frac{\bar{\Lambda}/\bar{\Sigma}}{\Lambda/\Sigma} =\frac{\bar{p}}{p} D;
\;\;\;\; \frac{\bar{\Xi}}{\Xi}=\frac{\bar{p}}{p} D^2;\;\;\;\;
\frac{\bar{\Omega}}{\Omega} =\frac{\bar{p}}{p} D^3  \label{3}
\end{equation}
where
\begin{equation}
D=\frac{q\bar{s}}{\bar{q}s}     \label{4}
\end{equation}
From these equations we see that the four ratios in (\ref{2}) and
(\ref{3}) are expressed in terms of two parameters. Therefore we have
two constraints which must be satisfied by the data. 

Let us first discuss the data for Pb-Pb collisions at CERN SPS.
The data of \cite{97}  give the following values 
for strange antibaryon/baryon ratios in the central rapidity region 
\begin{equation}
\frac{\bar{\Lambda}/\bar{\Sigma}}{\Lambda/\Sigma} = .133\pm .007;
\;\;\;\; \frac{\bar{\Xi}}{\Xi}=.249\pm .019;\;\;\;\;
\frac{\bar{\Omega}}{\Omega} = .383\pm .081 \label{5}
\end{equation}
The data of NA44 \cite{44} give 
\begin{equation}
\frac{\bar{p}}{p} = .07 \pm .01  \label{6}
\end{equation}
Dividing the ratios (\ref{5}) by the ratio (\ref{6}) and using (\ref{3})
we have
\begin{equation}
D_{\Lambda} = 1.9\pm .3 ;\;\;\;\; D_{\Xi} =1.89\pm.15;\;\;\;\;
D_{\Omega} = 1.76 \pm .15   \label{7}
\end{equation}
and thus we see that the three values of the parameter $D$ obtained from
the data are in good agreement with each other up the experimental
accuracy of about 10 percent.

From (\ref{6}) we also deduce that 
\begin{equation}
\frac{\bar{q}}{q} = .41 \pm .02   \label{8} 
\end{equation}
and thus employing (\ref{7})
\begin{equation}
\frac{\bar{s}}{s} = .75 \pm .06         \label{9}
\end{equation}
where we have used the average of the three values for $D$ given in
(\ref{7}), i.e. $D=1.83 \pm .10$. The ratios  (\ref{8}) and
(\ref{9}) are in good agreement with those obtained in \cite{let2} from
a thermal fit to the data \footnote{These ratios can be calculated as 
 the inverse
square of the corresponding fugacities given in \cite{let2}.}

This completes the analysis of the Pb-Pb data. Let us now turn to the
p-Pb collisions.

The data of WA97 coll. \cite{97} give 
\begin{equation}
\frac{\bar{\Lambda}/\bar{\Sigma}}{\Lambda/\Sigma} = .20\pm .03;
\;\;\;\; \frac{\bar{\Xi}}{\Xi}=.33 \pm .03;\;\;\;\;  \label{10}
\end{equation}
The $\bar{\Omega}/\Omega$ is not given in \cite{97}.
  The $\bar{p}/p$ ratio was
measured by NA44 collaboration \cite{441}, with the result
\begin{equation}
\frac{\bar{p}}{p} = .31\pm.03     \label{11}
\end{equation}
Using these values and the formulae (\ref{2}),(\ref{3}) we thus obtain
\begin{equation}
D_{\Lambda} = .65 \pm .11 ;\;\;\;\; D_{\Xi} =1.03\pm.07 \label{12}
\end{equation}
in clear disagreement. We must conclude that the quark counting rule is
apparently in contradiction with p-Pb data, indicating that in this case the 
quark
degrees of freedom do not represent a decisive factor in the production
mechanism.

Taken together, Eqs (\ref{7}) and (\ref{12}) show that the result
obtained for Pb-Pb collisions is likely not accidental but indeed
indicates existence of an intermediate step in baryon (antibaryon)
production process: a system of quasi-free constituent quarks and
antiquarks\footnote{Dynamical models which explicitely introduce the
$q-\bar{q}$ intermediate system  are being
developped since some time by the groups in Budapest  \cite{zi} and in Bratislava
\cite{pi}.} distributed
randomly in phase-space. A natural interpretation seems to be that this
intermediate $q-\bar{q}$ system is the first step of the chiral symmetry
breaking transition from the earlier quark-gluon plasma phase.

Another  interesting issue is: which  pattern is 
followed in collisions of lighter nuclei.
Answering  this question could bring new arguments to the controversy
as to where the transition to the quark-gluon plasma phase takes place.

The data  of \cite{441} and
\cite{ve} on $S-S$ collisions give:
\begin{equation}
\frac{\bar{p}}{p} = .12 \pm .01;\;\;\;\;
\frac{\bar{\Lambda}/\bar{\Sigma}}{\Lambda/\Sigma} = .22\pm .01;
\;\;\;\; \frac{\bar{\Xi}}{\Xi}=.55 \pm .07.  \label{13}
\end{equation}
So that we obtain
\begin{equation}
D_{\Lambda} = 1.83 \pm .17 ;\;\;\;\; D_{\Xi} =2.14\pm.16 \label{14}
\end{equation}
Thus the agreement with Eqs (\ref{2}) and (\ref{3}) is not bad, although
not as good as in the case of Pb-Pb collisions.

Using the average value $D=1.99\pm .12$ and $\bar{q}/q= .49\pm.02$ we
obtain $\bar{s}/s= .98\pm .07$ in good agreement with the analysis of
\cite{sol} where the data on central rapidity region in $S-S$ collisions
were discussed using the thermal model.
 It is also interesting
to note that the obtained value of the parameter $D$ is not inconsistent with
that found from the Pb-Pb data. This certainly supports the idea that
already in $S-S$ collisions the baryon and antibaryon production process
proceeds through an intermediate random $q-\bar{q}$ system. This
observation, in turn, strenghtens the evidence for quark-gluon plasma
phase present already in collisions of light nuclei.

We would like to close this paper with the following comments.

(i) Although we consider only baryon and antibaryon production, it is
tempting to extend the argument also to $K$ and $\bar{K}$ production.
Taking into account (\ref{4}), we obtain
 \begin{equation}
\frac{K}{\bar{K}} = \frac{q\bar{s}}{s\bar{q}} \equiv D. \label{15}
\end{equation}
The data \cite{ve} and \cite{49} give 
\begin{equation} 
\left(\frac{K}{\bar{K}}\right)_{S-S} = 1.91\pm.37;\;\;\;\;
\left( \frac{K}{\bar{K}}\right)_{Pp-Pb} \approx 1.8\; (no\; error\; given) \label{16}
\end{equation}

We see that these results are not in disagreement with the values of $D$
found from baryon-antibaryon data in $S-S$ and $Pb-Pb$
collisions\footnote{I could not find the data for $p-Pb$ collisions. The
$p-S$ data \cite{ve} give $K^+/K^-= 2.02\pm .14$ is strong disagreement
both values of D in (\ref{12})}.

(ii) The advantage of our argument is that the Eqs (\ref{2}) - (\ref{4})
can be applied to any phase space region (of course the specific
parameters may depend on the region). When the appropriate data are
available, it shall be thus possible to test the relevance of the quark
degrees of freedom also outside the central rapidity region considered
here. One may hope in this way to determine the kinematic domain where
the particles are dominantly produced by the intermediate step of
quark-gluon plasma. Furthermore, this should allow to determine the
rapidity dependence of the basic ratio $\bar{s}/s$. This last point is
particularly interesting in view of the recent suggestion by Letessier
and Rafelski \cite{let3} that the observed deviation of $\bar{s}/s$ from
unity is a reflection of the Coulomb interactions.

(iii) Our argument assumes  that the production
of all baryons and antibaryons in the central rapidity region of heavy
ion collisions proceeds by a common mechanism, i.e. "coalescence" of the
independently distributed quarks and antiquarks. Recent analysis of the
thermal model by Letessier and Rafelski \cite{let3}, based on the
$Pb-Pb$ data extrapolated to full phase-space, (and including Coulomb
corrections which modify somewhat the relations (\ref{2})-(\ref{4})),
indicates that the conditions for production of $\Omega$ and
$\bar{\Omega}$ may differ from those of other baryons. This is certainly
a serious possibility. The present experimental accuracy does not yet
allow, however, to draw definite conclusions about this problem.

(iv) We would like to repeat that the relations (\ref{2})-(\ref{4}) are
independent of the assumption of thermal equilibrium. When thermal
equilibrium is additionally assumed, one may produce many more specific
predictions, as discussed in detail in \cite{let,sol}. In particular, it
is possible to calculate the ratios of the rates of particles with
different strangenes content, a task which is clearly beyond the scope
of the present investigation.
 We feel, however, that our simple argument can
still serve a useful purpose of convincing a layman that the quark
degrees of freedom are essential for a correct description of particle
production in heavy ion collisions.

\vspace{0.3cm}
{\bf Acknowledgements}
\vspace{0.3cm}

I would like to thank B.Muller for raising my interest in the subject.
The discussions with F.Beccatini, R.Caliandro, R.Lietava, E.Quercigh,
J.Rafelski and K.Zalewski are highly appreciated. Thanks are also due to
M.Morando and E.Quercigh for a very kind hospitality at the Padova
meeting on Strangeness in Quark Matter, where part of this work was done.
This investigation was supported in part by the KBN Grant No 2 P03B 086
14.

\end{document}